\documentclass[pra,twocolumn,showpacs]{revtex4}
\usepackage{graphicx}
\usepackage{color}
\usepackage{amsmath}
\usepackage{tabularx,units}
\begin{document}

\title{Density wave patterns for fermionic dipolar molecules on a square optical lattice: Mean-field-theory analysis }
\author{K. Mikelsons and J. K. Freericks}
\affiliation{Department of Physics, Georgetown University, Washington, DC, 20057 USA}

\date{\today}

\begin{abstract}
We model a system of ultracold fermionic dipolar molecules on a two-dimensional square lattice. 
Assuming that the molecules are in their nondegenerate hyperfine ground state, 
and that the dipole moment is polarized perpendicular to the plane 
(as in the recent experiments on $^{40}$K-$^{87}$Rb molecules), 
we approximate these molecules as spinless fermions 
with long range repulsive dipolar interactions. 
We use mean field theory to obtain the restricted phase diagram as a function of the filling, 
the strength of interaction and the temperature. 
We find a number of ordered density wave phases in the system, 
as well as phase separation between these phases.
A Monte Carlo analysis shows that the higher-period phases 
are usually suppressed in the exact solution.
\end{abstract}

\maketitle


\section{Introduction} 
Recent experimental work~\cite{ni08} in ultracold molecule physics
has formed fermionic dipolar molecules of $^{40}$K and $^{87}$Rb in their
rovibrational and hyperfine ground state.  Quantum degenerate dipolar systems 
have long been sought after because their long-range interactions makes them interesting
for quantum computing applications and for understanding ordered phases that can arise from long range effects.
These systems are both complicated, but tunable, because the long range, anisotropic dipolar interaction
can be manipulated with an external electric field~\cite{zoeller},
and should give rise to rich physics with variety of different ordered phases of matter~\cite{lin10}.

The dipolar interaction is attractive 
if the dipole moments are aligned head to tail,
and this can lead to undesirable recombination effects if the molecules have an exothermic reaction
possible, as with ${}^{40}\textrm{K}$-${}^{87}\textrm{Rb}$ molecules \cite{ye_ultracoldchem}. 
This can be mitigated by confining molecules to a narrow two-dimensional layer,
and setting a strong electric field perpendicular to this layer,
so that the dipole moments of molecules are aligned with 
the external field and always align head-to-head and tail-to-tail.
In this case, the inter-molecule interactions are always repulsive.  
Note that we assume we can create a single isolated plane of dipoles, 
so there is no possibility to form superfluid pairing between planes, 
where the interactions are attractive again~\cite{halperin}.

We assume the dipolar molecules are loaded onto a single
square optical lattice with the molecules fully polarized perpendicular to the plane.
Since the ${}^{40}\textrm{K}$-${}^{87}\textrm{Rb}$ molecules 
can be prepared in their lowest hyperfine nuclear spin state, 
the internal spin degree of freedom is effectively frozen out, 
and the molecules can be described as spinless fermions.

Previous studies of spinless fermion models~\cite{uhrig93,halvorsen94,uhrig95} 
mostly focused on short range interactions and used the simplification 
of going to the limit of large spatial dimensions. 
These works show a rich phase diagram with 
charge-density-wave ordering and phase separation. 
There also has been some work in two dimensions focused on stripe physics~\cite{henley}.

More recent work, carried out in the context of 
ultracold dipolar molecules~\cite{yamaguchi10,sun10,lin10,carr10},
focused primarily on finding new exotic phases (liquid crystal, smectic, or nematic),
often achieved by tuning the direction of the external field 
with respect to the plane of molecules,
but has not elucidated the details of the full phase diagram
for the density-wave phases.

The paper is organized as follows:
First, in Sec.~\ref{sec:formalism} we introduce the model 
and describe the possible density-wave orderings. 
Sec.~\ref{sec:method} provides the details of the mean-field-theory formalism
 and the numerical calculations. 
The results are presented in Sec.~\ref{sec:results}, 
and they are further discussed in Sec.~\ref{sec:discuss},
which are followed by the conclusions in Sec.~\ref{sec:conclude}.


\section{Model}
\label{sec:formalism}
We describe the rovibrational ground-state dipolar molecules 
in their lowest hyperfine state with the following Hamiltonian:
\begin{equation}
H=H_k+H_p= -t \sum_{\langle ij\rangle} \left( c_{i}^{\dagger}c_{j}^{\phantom{\dagger}} 
 + h.c.\right) + \sum_{i \neq j} U_{ij} n_i n_j\,,
\label{eq:H_spinless_dipoles}
\end{equation}
where $c_{i}^{\dagger}(c_{i})$ is the fermion creation (annihilation)
operator for a fermionic dipolar molecule at site  ${i}$, 
$n_{i} = c_{i}^{\dagger}c_{i}$ is the number operator, $t$ is the hopping amplitude
between the adjacent sites, and $U_{ij} = U/|\vec{r}_i - \vec{r}_j|^3$ 
is the long-range dipole interaction,
with $\vec{r}_i$ the position vector for the site $i$. 
The lattice spacing is taken to be equal to one.
With this model, we have assumed that the hopping occurs 
only between nearest-neighbor sites, and that the molecules are always 
in the lowest level of the periodic optical lattice potential
(this second assumption is not always true when the interaction 
between the molecules is too large).
We have also assumed that the form for the dipole interaction is valid 
even when the distance between molecules is only one lattice spacing.

Due to the bipartite nature of the square lattice, 
the Hamiltonian can be rewritten in a particle-hole symmetric form and 
one can restrict to fillings that satisfy $f = \langle n \rangle  \leq 0.5$.

\begin{figure}[h!]
\begin{tabularx}{\textwidth}{m{1.6in} m{1.6in} }
\renewcommand*\arraystretch{1.25}
\begin{tabular}{ c c c c }
$N_c$ & identifier & $\vec{a}_1$ & $\vec{a}_2$  \\ \hline
1 & 1A & ( 1, 0) & ( 0, 1) \\ \hline 
2 & 2A & ( 2, 0) & ( 0, 1) \\
2 & 2B & ( 1,-1) & ( 1, 1) \\ \hline 
3 & 3A & ( 3, 0) & ( 0, 1) \\ 
3 & 3B & ( 2,-1) & ( 1, 1) \\ \hline
4 & 4A & ( 4, 0) & ( 0, 1) \\
4 & 4B & ( 2,-2) & ( 1, 1) \\
4 & 4C & ( 2, 0) & ( 1, 2) \\
4 & 4D & ( 2, 0) & ( 0, 2) \\ \hline
5 & 5A & ( 5, 0) & ( 0, 1) \\
5 & 5B & ( 3,-2) & ( 1, 1) \\
5 & 5C & ( 2,-1) & ( 1, 2) \\ \hline
\end{tabular} & \includegraphics*[width=1.55in]{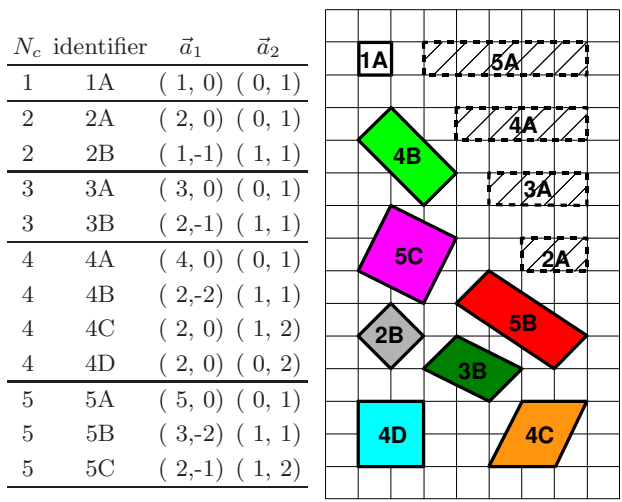} \\
\end{tabularx}
\caption{\label{tab:clusters} Left: Fundamental translation vectors 
($\vec{a}_1, \vec{a}_2$) for possible density wave phases considered in this work, 
grouped by the area of the unit cell. Unit cell 1A describes a
 homogeneous system without order, while 2B corresponds to checkerboard order. 
Right: Unit cells for the density wave phases. 
Only density wave orders corresponding to unit cells 
with the solid outline were found to be stabilized in this study.}
\end{figure}

Since there is no spin degree of freedom in this model, it can only support 
molecule density wave spatial order. Such order is 
described by its unit cell---a parallelogram spanned by the 
two fundamental translation vectors for a given density wave order.
The area of the unit cell ($N_c$) gives the number of sites
with independent average molecule density which is modulated 
periodically as the unit cell is tiled across the lattice. 
The number of the independent order parameters for this density wave is equal to $N_c-1$ 
(the average density, or filling, is not counted as an order parameter, 
and is fixed in the calculation).
Since there is only a finite number of non-equivalent unit cells 
of a given area on a square lattice, 
there is only a finite number of possible density wave orders. 
We have considered all such orders that have up to four 
different order parameters ($N_c\leq5$) and hence this is a restricted 
phase diagram restricted to low-period phases only.
The corresponding fundamental translation vectors and 
the unit cells for the twelve candidate phases are shown in Fig.~\ref{tab:clusters}.


\section{Methodology}
\label{sec:method}
We solve the model using mean field theory (MFT).
This can be justified, since the the interaction is 
long range and consequently each site is effectively coupled to any other site, 
thereby increasing the effective dimension of the mean field and 
decreasing the role of fluctuations (nevertheless, this is only an approximate solution).
In fact, due to the absence of a local interaction, 
the MFT is equivalent to the dynamical mean-field theory (DMFT) approach,
 which becomes exact in the infinite-dimensional limit. 
The absence of a spin degree of freedom also implies that 
the model is in the Ising universality class,
with a finite transition temperature in 2D. Thus, while MFT is expected to 
overestimate the transition temperature, 
it is also expected to give the qualitatively correct phase diagram.
Within MFT, the interaction part of Hamiltonian is approximated as:
\begin{equation}
n_i n_j \approx n_i \langle n_j \rangle + \langle n_i \rangle n_j 
 - \langle n_i \rangle \langle n_j \rangle \,,
\label{eq:MFT_approx}
\end{equation}
which means that the correlations of the density fluctuations are neglected.
This corresponds to the first order (Hartree-Fock) self-consistent 
perturbation theory result and is expected to be accurate for small $U/t$.

\begin{figure}[h!]
\begin{center}
\includegraphics*[width=3.0in]{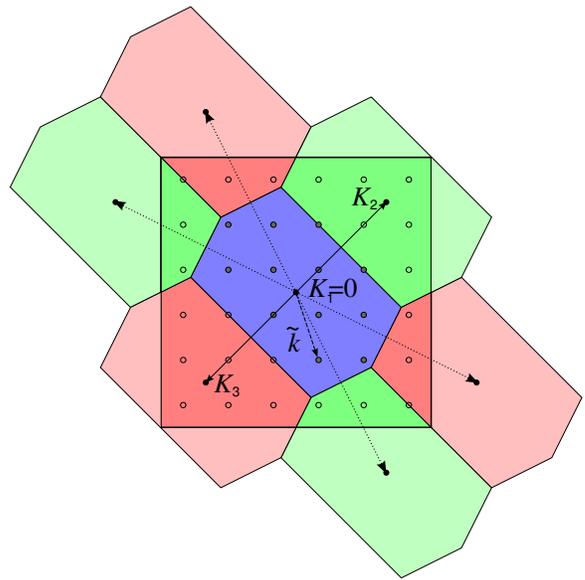}
\caption{(color online) Reciprocal space for the 3B order.
 The FBZ is shown in blue in the center of square.
The black circles denote $K$ points ($N_c=3$), associated with the density wave order, 
while the gray circles are the discretization $\tilde{k}$-points used in the calculation. 
The open circles denote other possible values of $K + \tilde{k}$ within reciprocal space.
The dotted arrows denote other reciprocal space vectors for 3B order 
used to construct the FBZ.}
\label{fig:3B}
\end{center}
\end{figure}

In the MFT approximation, the order parameter $\langle n_i \rangle$ is 
a fixed parameter in Hamiltonian, and acts as a site dependent potential. 
The resultant MFT Hamiltonian is quadratic in the ($c, c^{\dag}$) operators and can 
be easily diagonalized. We generically work on a lattice with a large, but finite, 
number of lattice sites $N$ and periodic boundary conditions. 
Further simplification comes from exploiting translational invariance,
which means that the Hamiltonian is block-diagonal in reciprocal space. 
The size of these blocks is $N_c$, and it grows with the number of the order parameters 
while the number of blocks, $N/N_c$, grows with 
the number of discretization points in reciprocal space, $N$, which is the same 
as the number of real-space lattice sites included in the calculation.
The Hamiltonian in reciprocal space becomes:
\begin{equation}
H = \sum_{\tilde{k}}^{N/N_c} \sum_{ij}^{N_c} c^{\dag}_{K_i + \tilde{k}} 
 \left[ (\epsilon(\tilde{k} + K_i) -\mu)\delta_{ij} + V_{ij} \right]  c_{K_j + \tilde{k}}\,,
\label{eq:H_MFT}
\end{equation}
where $\epsilon(k) = -2t(\cos k_x + \cos k_y)$ is the tight-binding dispersion, 
 $\mu$ is the chemical potential, $K_i$ are the reciprocal lattice points corresponding to 
the real-space basis vectors of a given order (see Fig.~\ref{tab:clusters}), 
and $\tilde{k}$ denotes the discrete summation points
 in the corresponding first Brillouin zone (FBZ) (see Fig.~\ref{fig:3B}).
The interaction part $V_{ij} = U(K_i-K_j) \langle m(K_i-K_j) \rangle$ where  
$m(K_i)$  is the Fourier transform of the real space density (order parameter):
$m(K_i) = \sum_{j=1}^{N_c} e^{i K_i r_j} n_j$, and $U(K_i)$ is:
\begin{equation}
U(K_i) = U \sum_{\textrm{all}\,j \neq 0} \frac{e^{i K_i r_j}}{|r_j|^3} \,.
\end{equation}
Since the the dipole interaction decays sufficiently fast with increasing distance,
the sum can be carried out for an infinite range of distance,
 giving more precise values for the effective interaction in $k$-space.
As seen from Eq.~\ref{eq:H_MFT}, $\tilde{k}$ is a good quantum number, 
while the spatially modulated mean field in the interaction term $V_{ij}$ 
causes molecule scattering among the $N_c$ $K$-points.
The calculation starts with a random guess for the order parameter 
$\langle n_i \rangle$ ($i=1,\ldots ,N_c-1$), 
which is put into the Hamiltonian (Eq.~\ref{eq:H_MFT}), 
and the full set of eigenvalues ($\epsilon_{\tilde{k},\alpha}$) and 
the corresponding eigenvectors [$\psi_{\tilde{k} \alpha}(K_i)$] are obtained
by solving the eigenvalue equation (for all $\tilde{k}$ and $\alpha = 1 \ldots N_c$): 
\begin{equation}
\sum_{j=1}^{N_c} \left[ (\epsilon(\tilde{k} + K_i) -\mu)\delta_{ij} + V_{ij} \right] 
  \psi_{\tilde{k} \alpha}(K_j) = \epsilon_{\tilde{k},\alpha} \psi_{\tilde{k} \alpha}(K_i)  \,.
\label{eq:H_MFT_eigenproblem}
\end{equation}
For simplicity, we denote $\epsilon_i = \epsilon_{\tilde{k} \alpha}$ and 
 $\psi_{ij} = \psi_{\tilde{k} \alpha}(K_j)$.
The energy per site and the filling fraction are next calculated via
\begin{equation}
E = \frac{1}{N} \sum_{i=1}^N \epsilon_i n_{\epsilon_i}\,\,\, \textrm{and} \,\,\, 
 f = \frac{1}{N} \sum_{i=1}^N n_{\epsilon_i}\,, 
\label{eq:energy_filling}
\end{equation}
respectively, with $n_{\epsilon_i} = 1/(1+e^{(\epsilon_i- \mu)/T})$ 
the Fermi-Dirac distribution and $T$ being the temperature. 
The mean-field entropy per site is
\begin{equation}
S = - \frac{1}{N} \sum_{i=1}^N \left[  n_{\epsilon_i} \ln ( n_{\epsilon_i} ) +
 (1 -  n_{\epsilon_i}) \ln (1 - n_{\epsilon_i}) \right]\,,
\label{eq:entropy}
\end{equation}
and the free energy satisfies $F=E-TS$. The momentum dependent density is:
\begin{equation}
\langle n_{\tilde{k} + K_l} \rangle  
= \langle c^{\dag}_{\tilde{k} + K_l} c_{\tilde{k} + K_l} \rangle 
= \frac{1}{N_c} \sum_{i=1}^{N_c}  n_{\epsilon_i} \psi_{il} \psi^{*}_{il} \,.
\label{eq:n_of_k}
\end{equation}
Calculating the expectation value of the average real space density
then yields the new values of the order parameter:
\begin{equation}
\langle m(K_j) \rangle = - \frac{1}{N} \sum_{i=1}^N \sum_{l,m=1}^{N_c} 
 n_{\epsilon_i} \psi_{il} \psi^{*}_{im} \delta(K_l-K_m-K_j)\,,
\label{eq:m_of_q}
\end{equation}
which can be used to calculate the structure factor $S(Q) = |m(Q)|^2$.
These updated order parameter values are inserted back into the Hamiltonian 
in Eq.~(\ref{eq:H_MFT}), and the iterative process is repeated until the order parameter
has converged to a fixed point. 
The chemical potential needs to be adjusted after every MFT iteration,
so as to maintain constant filling.

Solving the MFT equations represents finding a 
global minimum for the free energy functional in a multidimensional space. 
Since there can be multiple distinct stable solutions
(corresponding to local minima in the free energy), the calculation needs to be repeated 
for several random initial choices of the order parameter, and only the 
solution corresponding to global free energy minimum should be taken. 
More starting points are necessary for higher order phases, since the 
parameter space of solutions is larger. For phases with only one order parameter (2B), 
the MFT solution can be found by a successive bisection method.
Overall, we found that with 20 to 50 initial random order parameters, in more than half of 
the cases, the calculation converged to the lowest free energy minimum. 
The number of required integration points $N$ depends on the value of $U/t$: 
for $U/t \rightarrow \infty$ taking $N=N_c$ is sufficient, while for 
small $U/t$ finer discretization is needed to account for the $k$-dependence of 
dispersion $\epsilon(k)$. 
For a given $U/t$, we started with a smaller number of integration points $N$,
 and increased it by a factor of four, until further increase
 did not produce a significant change in the phase diagram.

The search for the MFT solutions is repeated for all possible density-wave orders
(as described above), and only the solution with the lowest free energy is ultimately taken.

A simplification occurs in the limit $U/t \rightarrow \infty$. 
In this case, the kinetic energy term in the Hamiltonian in 
Eq.~(\ref{eq:H_spinless_dipoles}) is negligible, and the MFT Hamiltonian becomes:
\begin{eqnarray}
H &=& \sum_{i \neq j} U_{ij} \left[ n_i \langle n_j \rangle 
 + \langle n_i \rangle n_j - \langle n_i \rangle \langle n_j \rangle \right] \\
  &=& \sum_i W_i n_i + const  \,,
\end{eqnarray}
with 
\begin{equation}
W_i = 2 \sum_{j \neq i} U_{ij} \langle n_j \rangle ~.
\label{eq:Wi_ni}
\end{equation}
The solution yields readily:
\begin{equation}
\langle n_i \rangle = \left( 1+e^{(W_i-\mu)/T} \right)^{-1}~.
\label{eq:ni_Wi}
\end{equation}
Equations (\ref{eq:Wi_ni}) and (\ref{eq:ni_Wi}) need to be solved 
self-consistently to yield the MFT result.

To account for the possibility of phase separation, 
we need to calculate the free energy for fixed $U/t$ and $T/t$
 for full range of filling $f$ and perform a Maxwell construction.
 A clear indicator of phase separation is that 
 the chemical potential dependence on the filling becomes multivalued
 (since $\mu = \partial F / \partial f$), as will be shown in the following Section.


\section{Results} 
\label{sec:results}
\begin{figure}[h]
\begin{center}
\includegraphics*[width=3.2in]{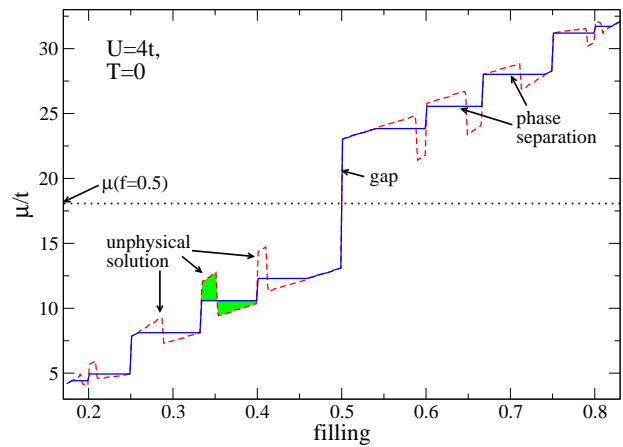}
\caption{(color online) Chemical potential for $U/t=4$ and $T=0$.
The red dashed line shows the MFT solution
 [which obeys particle-hole symmetry, $\mu(f)+\mu(1-f)=2\mu(0.5)$],
 but includes regions of unphysical behavior with $\partial \mu/ \partial f < 0$.
The blue line shows the physical solution, 
which includes regions of phase separation with $\partial \mu/ \partial f = 0$,
 accompanied by corresponding gaps at commensurate fillings.
 Two shaded regions near $f=0.35$ have the same area, according to Maxwell construction.}
\label{fig:phase_sep_U4}
\end{center}
\end{figure}

We present results for the phase diagrams, obtained both with and without 
phase separation considered. 
 To illustrate the procedure for detecting phase separation, 
 we show results for the chemical potential
 at $T=0$ and $U/t=4$ in Fig.~\ref{fig:phase_sep_U4}. The MFT solution 
 is particle-hole symmetric, so that $\mu(f) + \mu(1-f) = 2\mu(0.5)$. 
 However, it shows unphysical behavior at some regions
 where the chemical potential decreases with increasing filling.
 The physical solution avoids such regions by allowing for phase separation, 
 which consists of mixtures of two different phases with the same chemical potential,
 at the same temperature and interaction $U/t$.
  As seen in Fig.~\ref{fig:phase_sep_U4},
 phase separation occurs at fillings slightly above or below the commensurate values, 
 and is accompanied by a gap in the density of states
 at the commensurate values of filling.

\begin{figure}[h!]
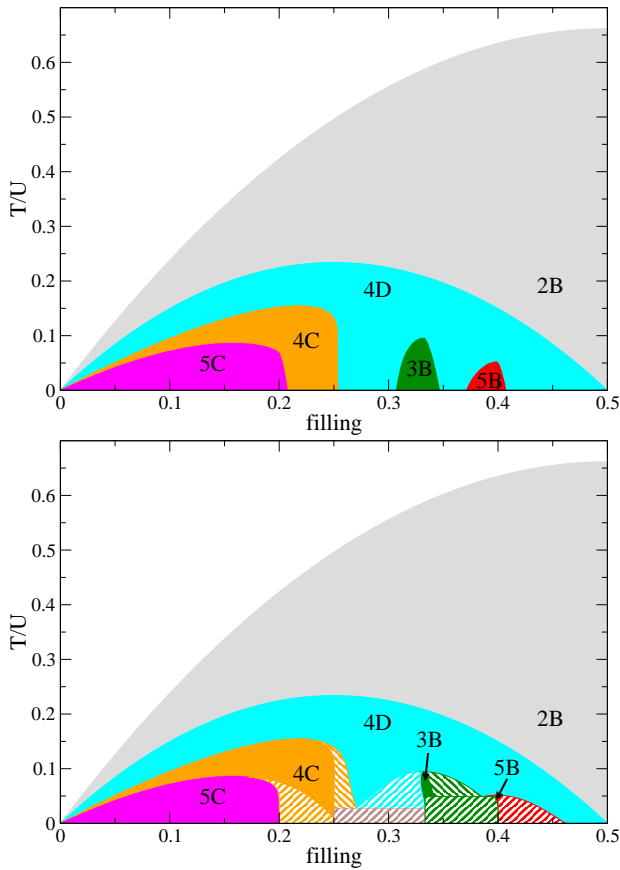

\begin{center}
\includegraphics*[width=3.2in]{FIG04_Tc_vs_fill_UInf.eps}
\includegraphics*[width=3.2in]{FIG05_Tc_vs_fill_UInf_PS.eps}
\caption{(color online) $U=\infty$ phase diagram 
without (top) and with (bottom) phase separation. 
Phase separation is shown as striped regions.
The transitions between 4D and 2B (as well as between 2B and 1A) phases 
are continuous. The rest of the transitions are discontinuous
 because of phase separation. }
\label{fig:phase_diag_U_oo}
\end{center}
\end{figure}

The phase diagram for $U/t=\infty$ is shown in Fig.~\ref{fig:phase_diag_U_oo}. 
In this case, the kinetic-energy part of the Hamiltonian is suppressed ($t=0$), and 
the order at zero temperature extends to all values of the filling. The dominant 
order is the checkerboard pattern (2B), commensurate at half filling. 
At low temperature, 
higher-period orders emerge around the corresponding commensurate fillings.
Yet low-period orders are more stable against thermal fluctuations, 
as indicated by their persistence up to higher temperatures.
The transition between the high temperature homogeneous (1A) to checkerboard phase (2B) is 
continuous, as is the transition between the 2B and 4D phases. 
The remaining phase transitions, which occur 
between higher-period phases at lower temperatures, are discontinuous and 
can involve phase separation.

\begin{figure}[h!]
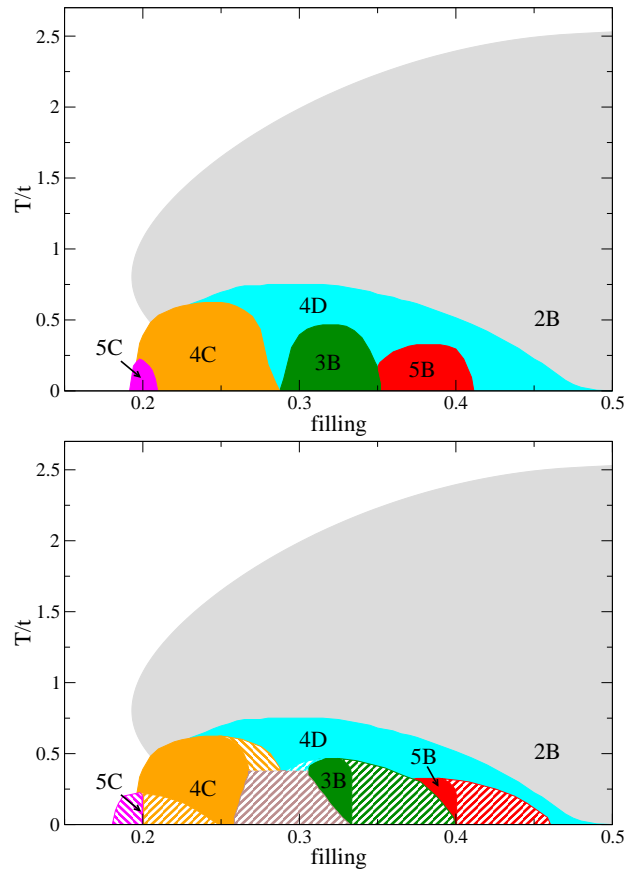

\begin{center}
\includegraphics*[width=3.2in]{FIG06_U4_phase_diag.eps}
\includegraphics*[width=3.2in]{FIG07_U4_phase_diag_PS.eps}
\caption{(color online) $U=4t$ phase diagram 
without (top) and with (bottom) phase separation. 
There is no order at low filling, 
yet for $f>0.25$, the phase diagram qualitatively resembles the $U/t=\infty$ result. 
Note the reentrant behavior for the checkerboard phase (2B) around $f=0.2$
and the phase separation between the 5C and homogeneous phases.}
\label{fig:phase_diag_U4t}
\end{center}
\end{figure}

For finite $U=4t$ (see Fig.~\ref{fig:phase_diag_U4t}), the zero-temperature order 
is suppressed for small values of filling, yet the phase diagram around half-filling
 is relatively unchanged, as compared to $U/t=\infty$.
 Again, the checkerboard phase is dominant at higher temperatures.
 Higher-period phases (3B, 4C and 5B), 
although still confined to low temperatures, appear to be more protracted in the range of filling.
The phase separation between these phases is also more prominent, but not qualitatively different
than in the $U/t=\infty$ case.
An interesting feature is the reentrant behavior for this phase around $f=0.2$
(evidenced by the backward curving of the stability region of the checkerboard phase). This is 
presumably induced by the van Hove singularity, which can stabilize the checkerboard order
 with the help of thermal activation. 

\begin{figure}
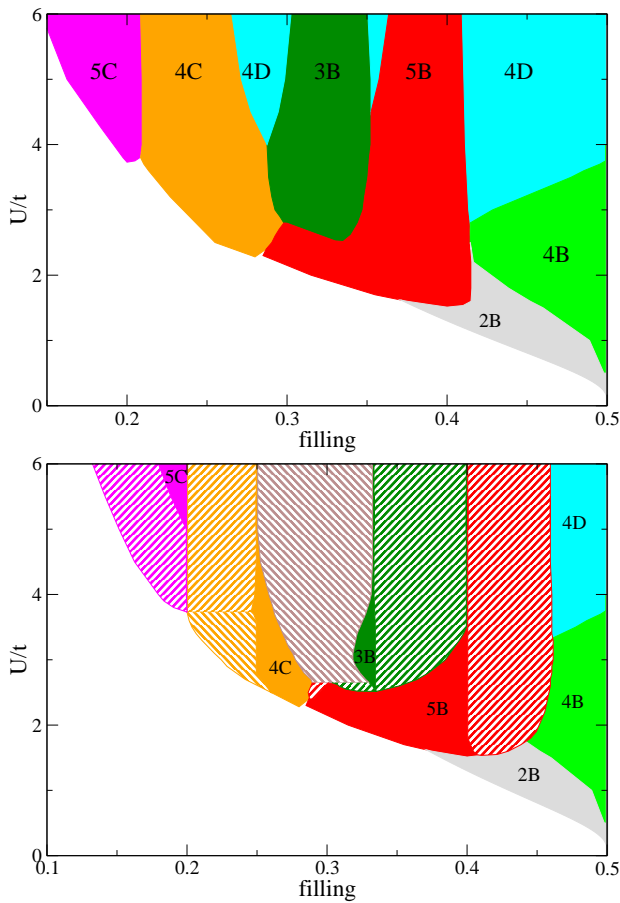

\begin{center}
\includegraphics*[width=3.2in]{FIG08_T0_phase_diag.eps}
\includegraphics*[width=3.2in]{FIG09_T0_phase_diag_PS.eps}
\caption{(color online) $T=0$ phase diagram
without (top) and with (bottom) phase separation.
 Decreasing the interaction
 contracts the range of filling of the ordered phases and progressively eliminates 
 phases commensurate with low values of filling. Only the checkerboard phase survives
 down to $U=0$. 
 Phase separation replaces the 4D phase 
 near $f=0.28$ and $f=0.36$ for larger $U/t$.
 In parts of the phase diagram, 4C and 5C phases show phase separation
 with the homogeneous state.}
\label{fig:phase_diag_T0}
\end{center}
\end{figure}

The phase diagram for $T=0$ is shown in Fig.~\ref{fig:phase_diag_T0}. 
Decreasing the interaction from $U/t=\infty$ causes ordering to disappear for low filling, 
as the kinetic energy effect is dominant in a dilute system. 
As the interaction is decreased, the phases with filling near commensurate values 
are cut off successively. Before its disappearance, 
phase 5B is significantly expanded in filling around $U/t=2.5$,
 perhaps due to a better nesting of Fermi surface, relative to other phases.
Below, but near half filling, decreasing interaction suppresses the 4D phase
in favor of weak order in the 4B phase and eventually
 leaving only the checkerboard order (2B).
The phase separation is not confined to the immediate vicinity of the commensurate fillings,
(also seen in Figs.~\ref{fig:phase_diag_U_oo} and \ref{fig:phase_diag_U4t}), 
as it can replace other phases for a range of fillings.

\begin{figure}
\begin{center}
\includegraphics*[width=3.2in]{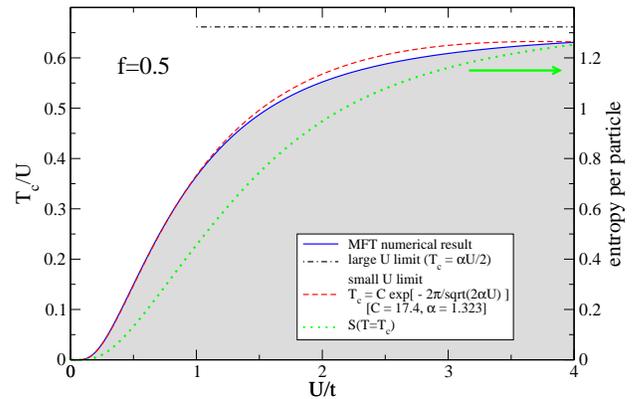}
\caption{(color online) $f=0.5$ phase diagram. For large $U/t$, 
 the transition temperature satisfies $T_c = \alpha U /2$,
 where $\alpha = U[K=(\pi,\pi)] = 1.323$. 
 For small $U/t$, $T_c$ fits perfectly to a semianalytical form.
The gap in the density of states (not shown) at $f=0.5$ and $T=0$ shows similar dependence on $U/t$.
The entropy per particle (dotted line) at $T=T_c$ reaches $S=2\ln2$ at $U/t=\infty$.
}
\label{fig:phase_diag_halffill}
\end{center}
\end{figure}

The checkerboard ordering is the most favorable phase at half filling
 for any value of $U/t$ and it persists down to $U/t=0$, due to perfect nesting
 of the Fermi surface (see Fig.~\ref{fig:phase_diag_halffill}).
For large $U/t$, the transition temperature $T_c$ is linear in $U$, the largest energy scale. 
(Since there is only one spin species, 
there is no energy scale associated with spin exchange, 
such as $J=4t^2/U$ in the Hubbard model.)
For small $U/t$, $T_c$ shows exponentially activated behavior
 and can be fit with a semi-analytical form, 
obtained by approximating the non-interacting density of states
with a logarithmic divergence representing the van Hove singularity. 
Similarly, we find an exponentially contracted range of filling
near $U=0$ (see Fig.~\ref{fig:phase_diag_T0}) for this phase.
Thus, checkerboard ordering near $U=0$, while possible due to perfect nesting, 
is strongly suppressed due to the proximity to the van Hove singularity. 
The gap in the density of states as well as the order parameter at $T=0$ show very similar 
dependence on $U/t$.

We also report the entropy per particle, since this parameter is important in experiments.
Calculated at $T_c$ for the checkerboard phase at $f=0.5$, 
the entropy per particle shows similar behavior with $U/t$ to that of $T_c$.
However, for $U/t=\infty$, the ordering starts at $S(T=T_c)= 2 \ln 2$,
clearly an overestimate due to the neglected density fluctuations in the MFT.
For smaller $U/t$, a lower entropy state has to be reached to observe the checkerboard order.

The most common experimental test for spatial order is  
Bragg diffraction of light by the molecules, 
which directly yields the structure factor in reciprocal space
and can be used to identify the order~\cite{hulet}. 
Such an experiment is the analog of an x-ray diffraction experiment 
on a condensed-matter system used to determine its crystal structure.
Table~\ref{tab:structure_factors} shows a comparison of different density-wave phases found in this study 
by listing the real space density patterns, reciprocal space points and structure factors for $U/t=4$. 
Each of the phases have sufficiently different signatures, both in the location and the relative strength of 
the peaks of $S(Q)$, making these phases easily identifiable through Bragg diffraction measurements.
\begin{table}
\begin{tabularx}{\linewidth}{m{0.55in} m{0.85in}  m{1.3in}  m{0.2in} m{0.2in}} 
phase & $k$-space points & structure factors & $f$ & $T/t$ \\ \hline       
\includegraphics[width=0.5in]{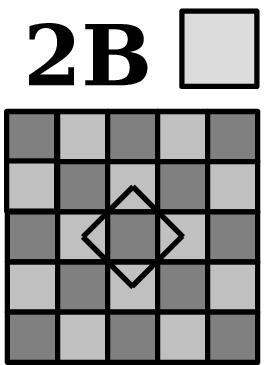} & \includegraphics[width=0.8in]{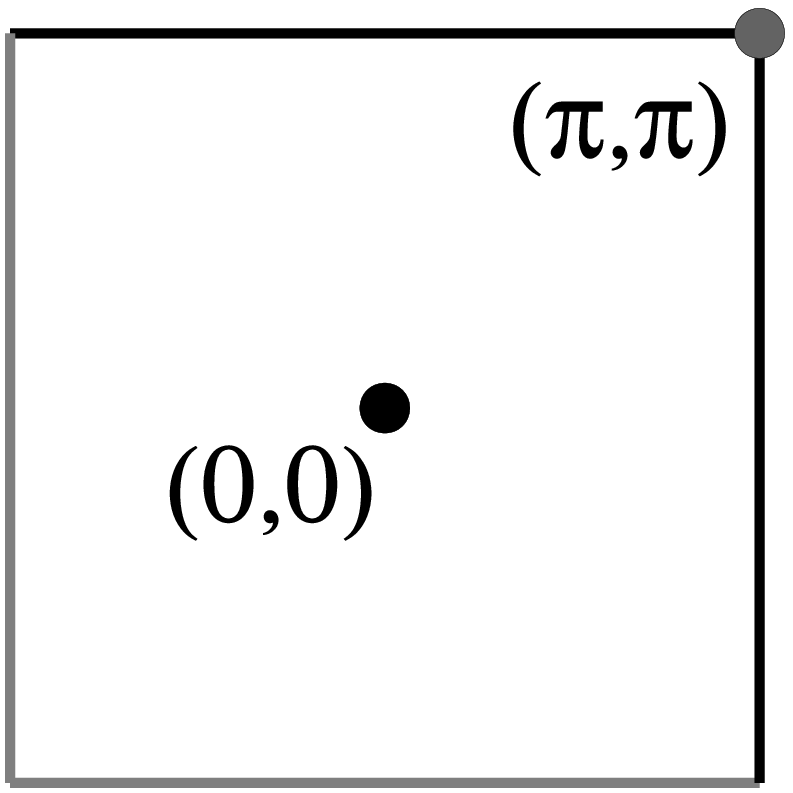} 
 & $S_{(\pi,\pi)} = 0.874 S_0 $ & \nicefrac{1}{2} & 0.1\\
\includegraphics[width=0.5in]{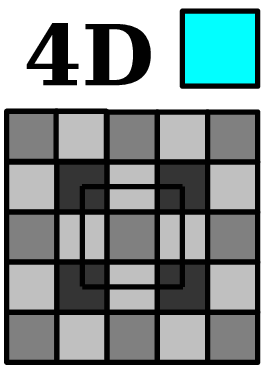} & \includegraphics[width=0.8in]{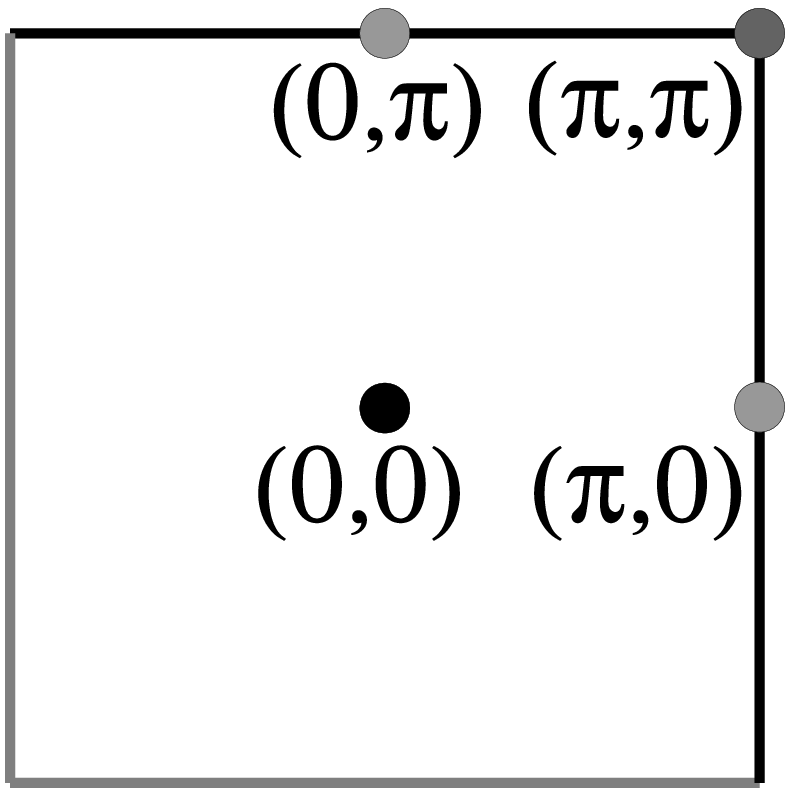}
 & \begin{tabular}{r c l} $S_{(\pi,\pi)}$ & = & $0.662S_0$ \\ $S_{(0,\pi)}$ &=& $0.279S_0$ \\
  $S_{(\pi,0)}$ &=& $S_{(0,\pi)}$ \end{tabular}& 0.3     & 0.5         \\
\includegraphics[width=0.5in]{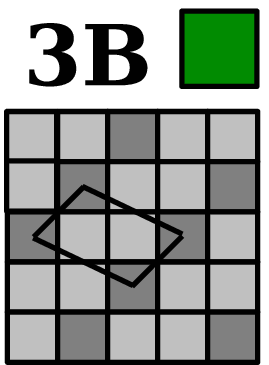} & \includegraphics[width=0.8in]{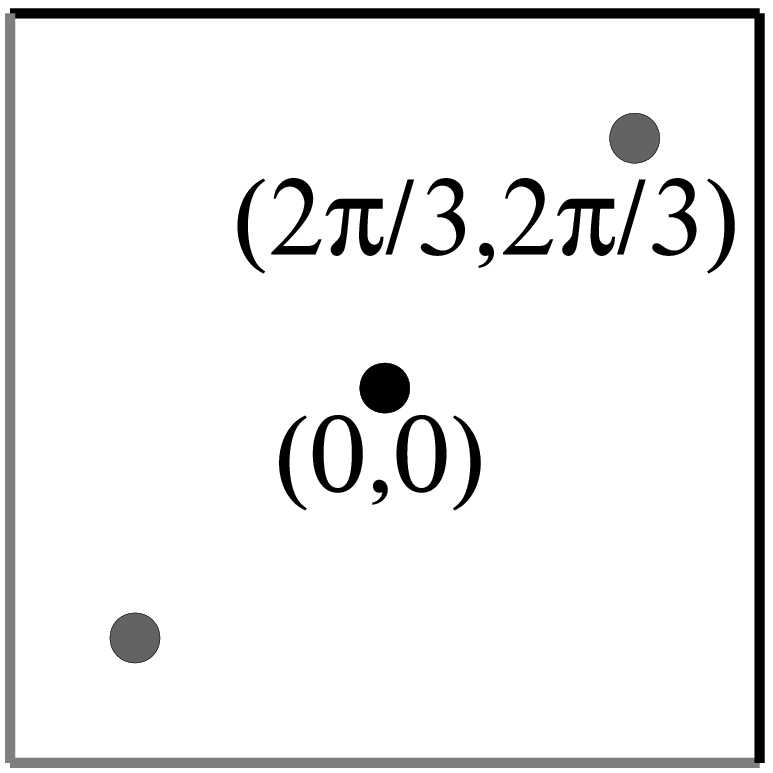}
 & $S_{(\frac{2\pi}{3},\frac{2\pi}{3})} = 0.751S_0$ & \nicefrac{1}{3} & 0.1 \\
\includegraphics[width=0.5in]{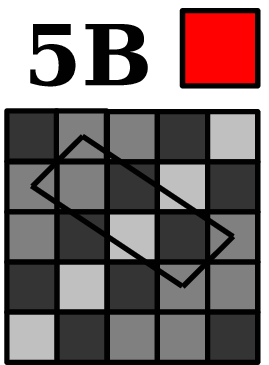} & \includegraphics[width=0.8in]{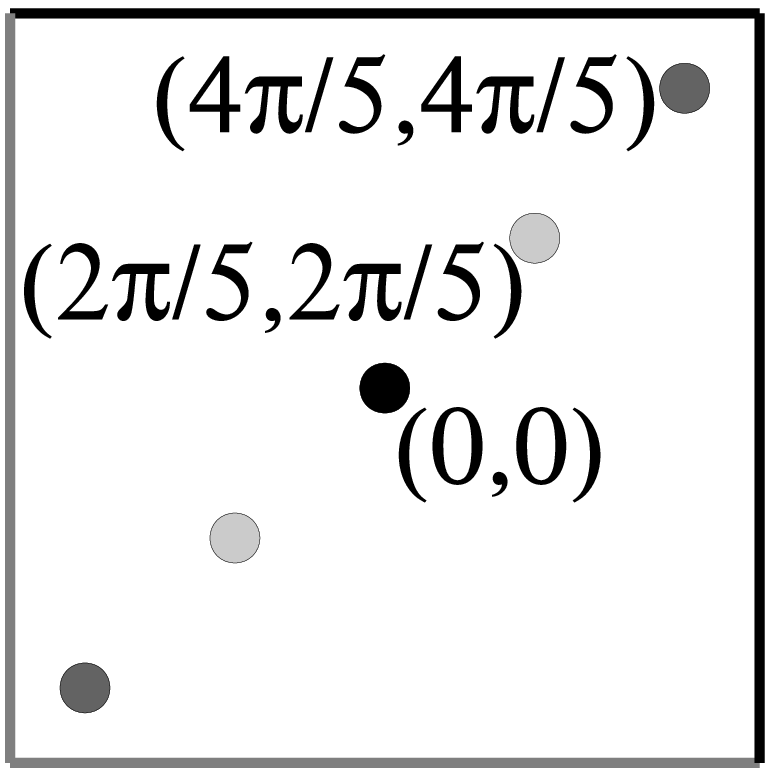}
 & \begin{tabular}{r c l} $S_{(\frac{2\pi}{5},\frac{2\pi}{5})}$ &=& $0.063S_0$ \\
 $S_{(\frac{4\pi}{5},\frac{4\pi}{5})}$ &=& $0.511S_0$  \end{tabular}  & \nicefrac{2}{5} & 0.1 \\ 
\includegraphics[width=0.5in]{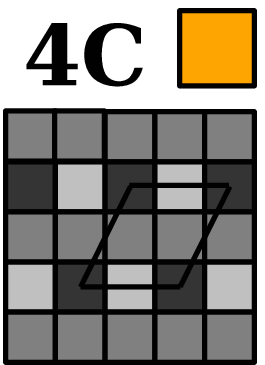} & \includegraphics[width=0.8in]{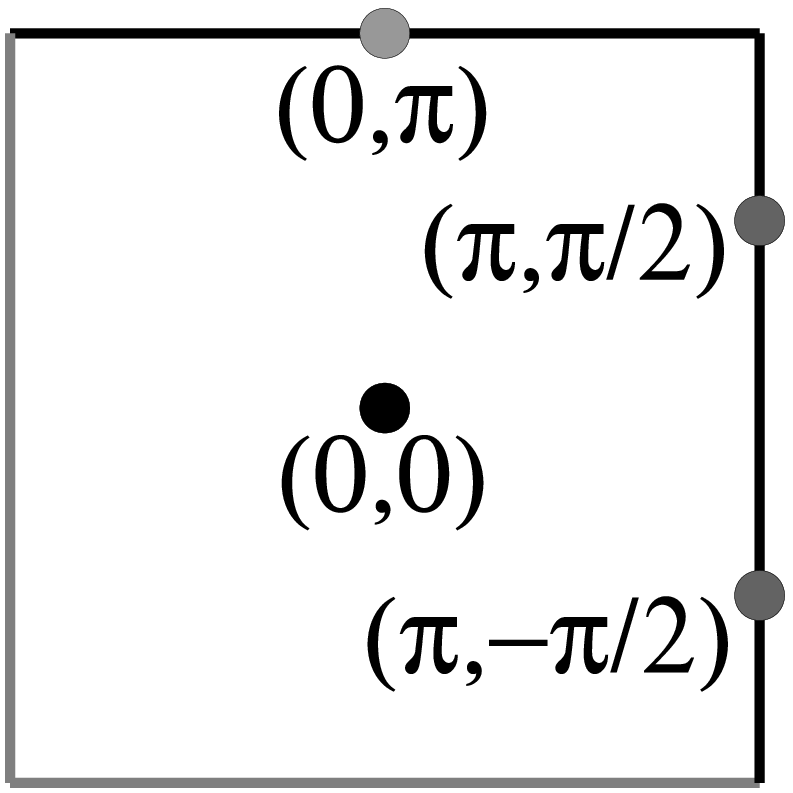}
 & \begin{tabular}{r c l}$S_{(0,\pi)}$ &=& $0.653S_0$ \\  $S_{(\pi,\frac{\pi}{2})}$ &=& $0.658S_0$ \\
 $S_{(\pi,\frac{-\pi}{2})}$ &=& $S_{(\pi,\frac{\pi}{2})}$ \end{tabular} & \nicefrac{1}{4} & 0.1 \\
\includegraphics[width=0.5in]{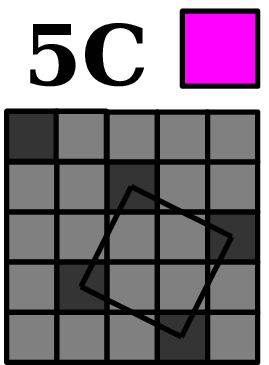} & \includegraphics[width=0.8in]{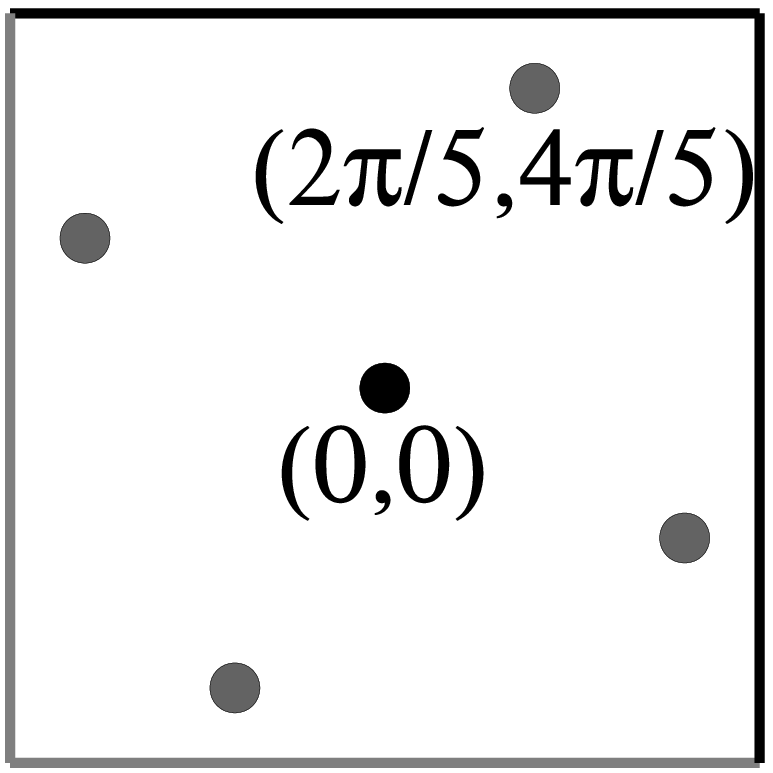}
 & \begin{tabular}{r c l} $S_{(\frac{2\pi}{5},\frac{4\pi}{5})}$ &=& $0.412 S_0$ \\
 $S_{(\frac{4\pi}{5},-\frac{2\pi}{5})}$ &=&  $S_{(\frac{2\pi}{5},\frac{4\pi}{5})}$\end{tabular}
 & \nicefrac{1}{5} & 0.1 \\ \hline
\end{tabularx}
\caption{Structure factors for different phases at representative values of filling and temperature for $U/t=4$. 
The first column shows the real space patterns with darker squares denoting larger average density per site.
The second column shows the reciprocal space points corresponding to a given density-wave order, followed  by
the corresponding structure factors at these points, $S(K)$, relative to $S_0 = S(K=0)$. 
Since $S(-K) = S(K)$, only the structure factors for $K_x \leq 0$ are shown.
The last two columns give the corresponding filling and temperature, respectively.}
\label{tab:structure_factors}
\end{table}

Another, less direct way to observe spatial order is time-of-flight measurements
that use the expansion of the molecular cloud to infer the initial momentum distribution
of the molecules. However, the momentum distribution data 
do not show prominent differences amongst the different ordered phases, 
so using this approach to detect the ordering is difficult. 


\section{Discussion}  
\label{sec:discuss}
The MFT results show a diverse phase diagram with a number of distinct density-wave phases.
However, the MFT is an approximate method, which neglects fluctuations and favors ordered states.
This brings to mind the question: 
{\it how many of these density-wave phases are actually present in the exact solution}? 

In the limit $U/t=\infty$, the Hamiltonian essentially describes a lattice gas model
 and can be solved numerically without approximation using classical Monte Carlo methods.
Preliminary results support the existence of the checkerboard (2B) and 4D phases, 
although strongly suppressed in temperature and filling, compared to the MFT result.
 Since the MFT neglects fluctuations, it is not surprising that
 the range of ordered phases is contracted in a more precise calculation.
 However, some higher-period phases become more prominent in the MFT solution 
 for intermediate values of $U/t$, so their presence cannot be entirely ruled out 
 by this $U/t=\infty$ study. A more complete study will be the topic of further research.

Also, the phase separation needs to be reviewed using better approximation techniques.
The presence of phase separation is tied to the existence of higher-period phases and the fact that 
the MFT strongly stabilizes these phases at the corresponding commensurate fillings.
Thus, phase separation is promoted by the MFT just like the corresponding ordered phases are.
It is possible that including even higher order phases
 (beyond 4 order parameters) in a MFT study 
 would replace the phase-separated regions with higher-period ordered phases
 that are commensurate at the corresponding values of filling, 
 yielding a devil's staircase-like structure. 
Another possible alternative to phase separation is the formation of a Wigner crystal like state.
The reentrant behavior of the checkerboard phase at $U=4t$ in the MFT results 
 is very likely an artifact of this method and should be reexamined using more precise techniques.

It is notable that the MFT results do not favor stripes that are aligned with the lattice
(corresponding to phases 2A, 3A, 4A or 5A). We found that these phases are 
either not stable or they are replaced by other phases with lower free energy. 
For small $U/t$, good nesting of the Fermi surface appears 
to be important to lower the energy of the system,
while for large $U/t$, strong repulsive interactions 
between adjacent sites prevents formation of such striped phases.
We also see no evidence for any nematic or smectic ``liquid-crystal'' phases.

Another important question is, 
{\it which of these density-wave phases can be reached and detected in experiment}?
The MFT results show large entropy per particle 
when the checkerboard state develops at half filling for large interaction, but 
unfortunately the entropy is severely overestimated in this study,
as shown by preliminary Monte-Carlo results.
The most easily observable features in experiment are the structure factors and 
 the gap in the density of states at the commensurate values of filling. 
These features are most pronounced for the checkerboard phase.
In a trap environment, a large gap at half filling 
should translate into large spatial regions of checkerboard order 
if the local density approximation is accurate.
Of course, it is not clear if phases that are seen in a homogeneous system
 are stabilized or destabilized by the trap environment. 
This is another question to be addressed in a future study.


\section{Conclusions}
\label{sec:conclude}
 We have studied the ordering of ultracold diatomic gases 
on a two-dimensional square optical lattice. 
We have formulated a model, based on spinless fermions
with long range dipole interactions, and solved it using mean field theory.
We have found a number of possible density-wave orders that occur mostly 
around the corresponding commensurate values of filling and 
create gaps in the density of states at these values of the filling.
The phase separation is also often seen between different ordered phases in the MFT results. 
Overall, the checkerboard phase is most stable near half filling 
and has the highest transition temperature. 
Due to the lack of an energy scale associated with spin exchange,
the transition temperatures increase with interaction strength.
Different density wave phases show characteristic and distinct structure factors.
We have discussed the validity of these results
 and the limitations of the MFT method used in this study.


\section{Acknowledgments} 
This work was supported by a MURI grant from 
Air Force Office of Scientific Research
numbered FA9559-09-1-0617  
and from a grant of HPC resources 
from the Arctic Region Supercomputing Center 
at the University of Alaska Fairbanks as part of the Department of Defense 
High Performance Computing Modernization Program.  
We also acknowedge support from the McDevitt endowment fund.


\end{document}